\documentclass{llncs}

\usepackage{amssymb,lscape}
\usepackage{graphicx}
\usepackage{enumitem}
\usepackage{multirow}
\usepackage{hhline}
\usepackage{url}
\usepackage{bbding}

\urldef{\mailsa}\path|lucas.gren@cse.gu.se|
\urldef{\mailsb}\path|magdalena.lindman@volvocars.com|
\urldef{\mailsc}\path||
\newcommand{\keywords}[1]{\par\addvspace\baselineskip
\noindent\keywordname\enspace\ignorespaces#1}

\begin{document}

\mainmatter  
\title{What an Agile Leader Does: \\The Group Dynamics Perspective}

\titlerunning{Group dynamics and agile leaders}

\author{Lucas Gren\inst{1,2} \and Magdalena Lindman\inst{2} 
}

\authorrunning{Lucas Gren and Magdalena Lindman}

\institute{Chalmers University of Technology and the University of Gothenburg, \\
Gothenburg, Sweden\\ 
\and
Volvo Cars,\\
Gothenburg, Sweden\\
\mailsa\\
\mailsb\\
\mailsc\\}

\toctitle{Lecture Notes in Computer Science}
\tocauthor{Authors' Instructions}
\maketitle

\begin{abstract}
When large industrial organizations change to (or start with) an agile approach to operations, managers and some employees are supposed to be ``agile leaders'' often without being given a clear definition of what that comprises when building agile teams. An inductive thematic analysis was used to investigate what 15 appointed leaders actually do and perceive as challenges regarding group dynamics working with an agile approach. \emph{Team maturity}, \emph{Team design}, and \emph{Culture and mindset} were all categories of challenges related to group dynamics that the practitioners face and manage in their work-life that are not explicitly mentioned in the more process-focused agile transformation frameworks. The results suggest that leader mitigation of these three aspects of group dynamics is essential to the success of an agile transformation. 
\end{abstract}

\section{Introduction}\label{sec:introduction}%what a dedicated agile leader does with regards to challenges!
The reported benefits of an agile approach includes increased customer collaboration, better estimation of tasks, and increased quality \cite{dybaa}, but also higher job satisfaction \cite{melnik2} as well as overall stakeholder satisfaction and, therefore, project success \cite{serrador2015does}. All of which have contributed to the popularity of the agile approach to development work. Agile development, compared to the plan-driven\slash waterfall, implies more communication and stronger focus on people, which make the social-psychological aspects very important to understand, including leadership and management \cite{lenbergchase}. 

%intro is not about challenges of agile leaders but about what agile leadership means and how teams also take on leadership. I would suggest to rather refer to literature on challenges that agile leaders face in the introduction 

The research on leadership was focused solely on the leader as an individual for many decades, often referred to as \emph{great person theory of leadership} \cite{hogg2003social}. The challenge for leaders was then to be categorized as a leader, but once accepted, the leader could probably act as he saw fit to a larger extent (leadership was also categorized as a male trait). The research on leadership and management then shifted from trying to find the best leader to looking at what the best leaders actually do, since it turned out not all the behaviors in one accepted leader were towards effectiveness in general \cite{hogg2014sp}. After that, instead of a focus on finding the best leadership style, a more systematic and situational view of leadership has developed \cite{2009tei}. What is the best leadership depends on the context, and, in some theories, on the maturity level of the co-workers, but it also contains a balance between task- and relation-focused behaviors connected to these phases \cite{hersey}. The leader role has thereby become more demanding and requires adaptability to context in a way that was not highlighted before. Trying to exert leadership from its context, the way researchers and practitioners have done, is probably a mistake. Schein \cite{scheincultlead} writes that leadership and culture are two sides of the same coin, and Northhouse \cite{northhouse} also emphasizes the importance of context, where he also models leadership as a collective ability of initiative. Dynamic team leadership \cite{2009tei} is not new in psychological science, and also a property of teams that are in the more mature stages of small group\slash team development \cite{tuckman}. 

Recent studies in software engineering has shown that the definition of agile teams overlap with what is meant by a mature team in social psychology \cite{grenjss2}. We define team maturity in this study as the degree to which a team has navigated through the group development stages according to Tuckman \cite{tuckman}. Teams transitioning towards agile ways of working are often in the middle of two paradigms where the more classical hierarchical management structure is in an organizational change process towards new forms of more dynamic and shared team leadership \cite{hodgson2013controlling}. Spiegler et al.\ \cite{spiegler2019leadership} showed that the leadership function is gradually transferred from the Scrum Master to the team members over time. In more general leadership science, Millikin et al.\ \cite{millikin2010self} showed that self-managing teams have higher productivity even in multi-team settings and that the highly cohesive teams perform the best. In the agile space, e.g.\ servant leadership is advocated as the foundation for leading, but the definition of what that is remains vague \cite{parris2013systematic}. It is, therefore, still unclear what the behavior and challenges are in relation to group\slash team dynamics when implementing agile teams from the perspective of an appointed leader. 

If the goal of leadership in an agile world is for the teams to eventually lead themselves within their mandate and technical expertise, we can also look at psychological theories on how to lead towards self-organization. We define self-organization in this paper as a state where the initiative, responsibility, and drive towards accomplishing team goals are dynamically shared between many team members in the same team. In past psychology research on self-organizing teams, team design has been found to be more important than team coaching when striving for self-organization \cite{wageman2001leaders}. In this context, Wageman \cite{wageman2001leaders} defines team design as including all the following eleven design features: (1) real team, (2) clear direction, (3) appropriate size, (4) skill diversity, (5) task interdependence, (6) challenging task objectives, (7) core strategy norms, (8) team excellence recognized\slash rewarded, (9) information for planning available, (10) training\slash technical consultation available, and (11) material resources available. We define team design in this paper as the eleven features above but want to stress that a role that is less integrated in the team compared to team members, needs to help design the team with regards to these eleven features. In our experience, agile teams are sometimes set up so that the teams themselves are expected do the design work, which is not what Wageman \cite{wageman2001leaders} found to work well. She instead highlights the external leader role in enabling teams to self-organize over time. 

Furthermore, Wageman \cite{wageman1997critical} describes the different roles of a leader as first having to be a designer, which includes (1) ``setting a direction for the performing unit, design a team task and a team reward system,'' (2) ``making sure the team has the basic material resources it needs to do the work,'' and (3) ``establishing the team's authority over and its responsibility for its performance strategies.'' Only later can an appointed leader be what is refer to as a midwife that should act at natural breaking-points in the team's further development, which comprises ``working with the team to establish appropriate performance goals.'' These goals should be measurable and specify ``how a team will take on its work in ways that fulfill its overall direction.'' Only in the final step should the leader be a coach. Team coaching is only possible with the other two steps fulfilled and only then will the team make good use of the coaching. She continues to state that ``because well-designed teams are robust enough to bounce back from inappropriate leader actions, the leader now has the latitude to unlearn old managerial habits and take the time that is needed to learn effective team coaching skills'' \cite{wageman1997critical}. 

We define leadership as a function of initiative or group action \cite{northhouse}, and therefore, view all the described behavior as components of leadership. Our research goal is to understand what appointed agile leaders do when building and maintaining agile teams and where that fits into related work on leadership behavior. A qualitative research method allows for a deeper analysis of the complexity of a construct, and allows research participants to speak freely about their reality. Therefore, a qualitative approach is appropriate to study leadership in the context of agile teams. This study aims at investigating how agile leaders at different positions at different industrial development departments interpret the, vaguely defined, concept of agile leadership in relation to group dynamics in their real-world context. The research question is, therefore, \textbf{RQ: What are the behaviors and challenges in relation to group\slash team dynamics when implementing agile teams from the perspective of an appointed leader?}

%Theory:  \cite{rising2000scrum}\cite{scrumorbeing}\cite{adkins}\cite{rising2000scrum cite{kalliamvakou2017makes}

%we do not adhere to any specific leadership theory, but instead want to describe how leader interpret their new role in an agile context. 
%torn between: 1 challenges of team dynamics, 2 what a dedicated leader does, 3 which kind of leadership the team takes over. Choose 2!!!!
%The authors should define the concepts they use: What is team design? What is team maturity? What does the term “self-organization” mean in this paper? And how it is different from leadership?
%we do not adhere to any specific leadership theory, but instead want to describe how leader interpret their new role in an agile context

\section{Method}\label{sec:methodology}
This section presents the method we used to analyze the leadership situation in the agile development context.

\subsection{Procedure}
The participants were obtained indirectly through our industry or research contacts. We asked these contacts to suggest fitting participants which we then contacted by email. A heterogeneous sample was achieved by recruiting people from many different companies, both with an overview of the development part of the organizations and people in new agile roles of newly formed teams. All of the people that we contacted participated. We conducted 45 to 90 minute open-ended interviews, and thirteen of them were conducted using teleconference. First, a personal introduction of the researcher(s) was done including research background, and what the researcher wanted to find out though the overall research project. The interviewer then asked for permission to record the interview and emphasized the anonymity of the data collection. Two interviews were conducted face-to-face and recorded on a mobile device, but were transcribed in the same way. Thirteen of the interviews were conducted by the first author in English, and two were conducted by the second author but in Swedish. The interviews were transcribed verbatim afterwards and Swedish quotes were translated into English.

\paragraph{Interview Protocol.}
Most of our questions were descriptive in nature, however, some were also contrasting and reflective. The reason why we wanted to be concrete and not ask directly about emotions and interpretations was that we wanted to meet the engineering at their own discourse, i.e.\ use vocabulary that engineers are used to in their work situation. If the interviewee expressed frustration, emotion or problematized something we asked follow-up questions to prone the person's interpretation and experience around that topic. 

The interviews were semi-structured and aimed to answer the research question on what main challenges agile leaders define in connection to group dynamics. We selected participant who saw themselves as leaders in the agile context an did not use any specific role or definition thereof. Examples of questions used to investigate such association were: ``What do you think is working\slash not working with the agile implementation and why?'' ``Do you see a difference in how high performing teams adopt agile compared to new or less mature teams?'' and ``How do work processes evolve in agile teams?''

\subsection{Participants}
The participants were practitioners working with an agile approach, according to themselves, on different levels of organizations, ranging from team Scrum Masters to founders or CEOs. The first thirteen participants were involved in software development and the second two were involved in hardware development complementing the sample since agile has spread to other areas than software development. However, the conclusions drawn from the hardware part should be considered with care since generalizations most likely cannot be drawn from only two participants, i.e.\ we have yet to see saturation in that data.

\begin{landscape}
\begin{table*}
\footnotesize
\renewcommand{\arraystretch}{1}
\caption{Company Information.}
\label{fig:companies2}
\begin{tabular}{p{20mm}p{53mm}p{41mm}p{41mm}}
\bfseries  & Interviewees \bfseries & Method used before agile & Reason for agile\\
\hline
Company $A$ & Agile coach  & Waterfall process &More engagement, job satisfaction, and quality.\\
Company $B$ & Project manager lead & Waterfall process & Improved business value.\\
Company $C$  & Project portfolio management responsible & Ad hoc process & Focus on project priorities instead of personal interest. \\
Company $D$ & Project Manager (initiated agile). & Waterfall process & Innovative ways to deliver value. \\
Company $E$  & Interviewee 1: Team leader sales and distribution. Interviewee 2: Lead of ~25 project managers & Waterfall process &Improve the company.\\
Company $F$ & Project manager in project execution & Waterfall process & Had read about agile methods, started a pilot project. \\
Company $G$  & Interviewee 1: Scrum Master/Project Manager. Interviewee 2: Scrum Master of two teams. Interviewee 3: Scrum Master  & Started as agile teams &Realized that the products needed to be developed faster. \\
Company $H$ & Scrum Master/manager in one of the first agile teams  & A culture of guessing what users liked & Better and faster feedback and solve organizational problem.s \\
Company $I$ & A multi-type supporting role & Something very similar to agile methods before they had heard about the concept.      &Started as an agile company.\\
Company $J$ & Founder of an agile company ($>$10 years experience) & A company built on agile principles and values in a very flat organizational structure. & Started as an agile company. \\
Company $K$ & Interviewee 1: Scrum Master driving the agile transformation of the team. Interviewee 2: Certified Scrum Master and developer & Waterfall process &Wanted to adapt to the fast development of new technologies \\
\hline
\end{tabular}
\end{table*}
\end{landscape}

Table~\ref{fig:companies2} provides a brief guide to the variety of cases that we investigated with information about the interviewees, and their organizational situation. The interviewees worked in companies ranging in size from 35 to 56 000 employees, and represented work cultures in seven different countries. Three of the participants where in an environment that started with an agile approach to work from the beginning. However, since they still described the agile way of working in contrast to their own and their colleagues previous ways of working, we opted to analyze all the transcripts in the same way.

\subsection{Reflexivity}
In accordance with Braun and Clarke \cite{braun2006using}, we believe that we ``cannot free ourselves of our theoretical and epistemological commitments,'' thus, we acknowledge our previous knowledge of group dynamics as researchers as well as experience from working in various teams and from leadership. We, therefore, acknowledge that we cannot be completely objective when interpreting the challenges stated by the participants. However, we did not have a preexisting coding scheme and tried to let the participants speak freely about their experiences without us intervening. The guidelines by Dahlberg et al.\ \cite{Dahlberg2008rlr} were applied in that a researcher must be prepared that data can present things differently than what was initially thought, i.e.\ we wanted to be surprised as researchers. The intention was to apply intellectual honesty, and thoroughness in reasoning and in view of condition and consequences. We then also want to avoid favoring one's own person, skewed sampling, omission of negative evidence, one-sided maneuvers and wishful thinking \cite{Dahlberg2008rlr}. The second author is an expert engineer without previous knowledge in agile methods but a lot of experience from product development, and could therefore provide an eyes-open-wide approach to the challenges studied. The first author is a researcher on the topic of building self-organizing agile teams and is therefore knowledgeable about the agile approach. However, he sees the agile methods as having a potential of positively transforming many parts of an enterprise, but also sees challenges with how it is sometimes implemented in practice. Therefore, the first author had no preconception about what results would be better or worse in conducting this study.

\subsection{Analysis}
We analyzed the transcripts using the six phases of thematic analysis suggested by Braun and Clarke \cite{braun2006using}. The first step was to read the entire transcripts before coding took place. We then consistently focused our analysis on statements regarding challenges of agile leadership in relation to group dynamics in concordance with our research question. The analysis was inductive in nature since we wanted to keep an open mind to emerging themes in the data. The first author began by coding the first thirteen interviews and the second author coded the remaining two. We then cross-checked two transcripts and themes and we discussed and resolved discrepancies. Then, as the third step, different codes were put together into themes and (the fourth step) checked whether we agreed upon which themes fit with the codes. We applied descriptive coding \cite{Saldana2015tcm} since our research question is in relation to finding separate challenges of group dynamics that the practitioners working with an agile approach need to manage in their daily work. Therefore, connections or hierarchies in the found challenges were not sought for. The fifth step was then to assess the naming of the challenges and, as a final step, connect them to existing literature. Our epistemological approach leans towards a phenomenological view more than the social constructionist one, since we believe the challenges of group dynamics in relation to leadership roles can be described directly. We believe this partly due to the fact that our interviewees deal with group dynamics in practice every workday, that these concept are less emotional compared to other psychological constructs. They were therefore expected to be able to articulate their challenges as agile leaders accordingly \cite{willig2013introducing}.

\section{Results}\label{sec:results}
The three main leadership challenges found in our data is summarized below. The challenges are shown below together with quotes and a discussion to support the claims.

\subsection{Team maturity}
When the interviewees compared less and more mature teams, the latter were said to tailor their own agile process based on contextual knowledge. One interviewee saw a strong connection between more mature teams that have met for a longer period of time and how much initiative and responsibility they take for the process and collaboration, implying that their leadership role is easier and more consultative. 

\begin{quote} 
``They reinforce the practices within the teams themselves.'' [Project manager and initiator of the agile approach]
\end{quote}

Another indication that the level of team maturity is highly influencing the appointed leader's leadership style in agile teams, is the fact that more mature teams were said completely adopt the agile practices they find useful, while less mature teams need reinforcement of the practices, otherwise they are reluctant to use them. To enforce the practices is then something the leaders describe that they must do for the team. 

\begin{quote} 
``Those are very visible high performing teams self-directive aspects vs.\ those needing reinforcement of the practices for them to be there.'' [Project manager and initiator of the agile approach]
\end{quote}

A key seems to be to suggest best practices by the leaders for the team, have a minimum of what is allowed, and then let the teams tailor their process themselves. In such a way, less mature teams will resort to safety by adopting a predefined process but can then redefine their process as the team matures.

\begin{quote} 
``We really try to get the teams to focus on staying within the framework, but they have latitude and liberty within that framework to, based on their own team style or team makeup.'' [Project Management Leader]
\end{quote}

Another interviewee clearly stated that even the most self-organized teams were different initially. The ``agility'' simply had to wait for the team to mature, which means that the needed leadership style is different across time. 

\begin{quote} 
``Yeah, these days they don't need me in order to work. These days I am really a facilitator and the team is absolutely able to the normal Scrum process without me. They don't need any guidance any longer so I can easily go on vacations for 2-3 weeks, that's not a problem. /.../ Of course in the beginning I had to stop them in the dailies and say `Stop discussing solutions,' just the tasks, please, and the three questions bla bla bla. Now it's more or less routine.'' [Scrum Master]
\end{quote}

This indicates that the self-organization of teams emerges over time along with the team maturity from a psychological perspective, which also implies that the leaders must take on the function of leadership initially before that function can be shared. In addition, the built in flexibility of the agile processes is also something teams need maturity in in order to leverage in the intended way. One interviewee stated that more mature teams can easily change their process if asked, which is not something the less mature teams could do in the same way. 

\begin{quote} 
``We are trying to adopt Kanban. But there are other teams here that have only worked with Kanban for a few months. They tried it because they saw some problems we had with Scrum. Some of the teams matured faster (mainly because they didn't break up the teams all the time as we do here) and they changed to Kanban.'' [Scrum Master]
\end{quote}

For teams that are mature, and where the members are dedicated and have set clear roles, the interviewees saw that they could adopt self-organization and team agility without many issues, putting less focus and dependence on them as the designated leader or manager. 

\begin{quote} 
``The team we have is an extraordinary team with a very open mindset and a very innovative team and always open to new things, so they were very open-minded so that was not much to say, and they trusted me.'' [Project Manager Lead]
\end{quote}

To summarize, the first found challenge is that the agile leader needs to take a step back from mature teams and instead facilitate the team's work process in relation to the surrounding ecosystems of the teams. However, in order to implement agility in the less mature or newly formed teams, they also need to provide a lot more direction and guidance in order for such teams to become self-organizing agile teams.

\subsection{Team design}
The team design process is very much connected to team maturity as described in the previous section. However, the following quotes are in this category because they are a symptom of a lack of team design from an external role, not a symptom of the team being new. 

When teams were not given an initial structure, some interviewees were surprised and frustrated when newly formed teams expressed a need for clear and directive leadership. 

\begin{quote} 
``It was the first retrospective that we had, they say that they lack some leadership there and then during the retrospective we were talking like `no it's not leadership that you need, how about that you decide how you will do things and the new habits that we will create in the next sprint.' What kind of agreements that we have to have for solving that problem, and we're saying that the lack of leadership is the result. That's one thing that appears a lot because we are growing and new teams are being formed.'' [Founder of an agile company]
\end{quote}

New team were also described as being open to any work practices and lack the insight into what is useful or needed in their context. This entails that the leader needs to step in and guide the team in making such decision, something that was described frustrating for the leaders since they did not expect that in the agile approach. However, the team must get help in its initial design since no team members can know the context simply because it is new to them.  

\begin{quote} %viktig
``Yeah, for teams that are younger, like formed more recently, they tend to be more open to all the practices, but they don't have the experience to decide which practices that would be the best.'' [Founder of an agile company]
\end{quote}

Also in relation to new teams, one interviewee changed the agile practice of volunteering for task because the team was not ready to take on that responsibility on a team level. This implies that the leader felt the need to step in and be more directive, however, with the expressed frustration that such a leader behavior is not appropriate in the agile context. Teams, though, seem to need help in designing work processes initially. 

\begin{quote} 
``In agile people should volunteer for tasks, but in most cases here we are obliged to… we do task assignment, by me or the person who already worked on this item takes this item. I know that this isn't a good practice in agile, but we do it for our team and for more productivity, but also for responsibility.'' [Scrum Master]
\end{quote}

An interviewee from hardware development also highlighted that, since the company is expanding all the time and therefore consistently gets more team members, it is difficult to design and build self-organizing teams, which is a core part of the agile approach and frustrating for them as leaders. 

\begin{quote} 
``You have to be very involved to be able to get a clear direction from it, and that's hard, especially because our group has grown so much and we get new people all the time who don't have that direction from the beginning. /.../ The further the team gets, and the more you have worked, the more autonomous the group becomes.'' [Scrum Master and Software Developer]
\end{quote}

We also found support for the distinction between what is an organizational and enforced structure and what is up to the teams themselves, i.e.\ the right balance in team design. Teams with no provided structure were described as much less effective but it is about providing the right balance of flexibility and control by the people in leadership positions. 

\begin{quote} 
``The teams can change as they want [in the process] all the time.'' [Supporting role]
\end{quote}

Traditionally, it seems like most software development processes did not have any team reflection sessions by default. The agile process often adds the retrospective meeting, which is a structure for team reflection, which was described as helpful for the leaders in order to improve the teamwork. 

\begin{quote} 
``The developers, I think, also feel that it provides them with a preset structure within which they can communicate with each other; they don't have to set up a meeting to do this. We have our Scrum identify that they need to meet to talk about something, and then they do. So it puts things in place for them and they don't have to think about it.'' [Project Management Leader]
\end{quote}

To summarize, the leaders need to help transitioning teams to design their new agile structure and ways of working. If teams are not ready to tailor their own process, the leaders provide suggestions and best practices, but keep stating that the team should continuously improve their process based on what they learn about their ecosystem. This means that, when the teams are set up, they need an initial predefined team design and then be given the possibility to tweak their processes when the team is ready.

\subsection{Culture and mindset}
The third and final challenge was that the leadership also needs to be adjusted to the existing more traditional structure and culture of the company. We found that even if members and teams can adopt self-organization, the context might hamper this way of working. 

\begin{quote} 
``If the whole company was agile in the end, the teams could be more independent and talk to the business area more themselves, but now I spend a lot of time trying to make decisions on what the teams should do.'' [Scrum Master]
\end{quote}

All the interviewees spoke about the different agile roles changing dynamically based on what is really needed at a point in time. A project manager acts as a Scrum Master, or trying to help people to not fall into old patterns of behavior in relation to the old structure instead of the new agile process.

\begin{quote} 
``After the first project I was able to spend time educating them at the beginning, but had to make almost daily conscious efforts at reminding them, or educating them with when I could see that their though process was tending toward waterfall. So I'll try to point out: `oh actually, let's think about it this way.' Or, you know, helping them with definitions like what the basic function of the daily Scrum for example and remind them of that. That it's not a status meeting, for example.'' [Scrum Master and Project Manager]
\end{quote}

One participant from the hardware-focused development also highlighted the fact that all individuals do not have the same possibilities, or motivation, to adapt to the new ways of working. The participants from hardware and software development differed in that the hardware-focused interviewees focused on the ``old'' ways of working as compared to the new agile way, while the software-focused participants reflected more on the team's place in the company as a whole.

\begin{quote} 
``Partially, I believe it's due to... from what I've heard, that [a team member] that is a bit more senior and has had previously bad experience from the agile ways of working in other areas and therefore doesn't think it worked well and actually only sees the negative aspects. [The team member] just does this because everybody else wants to do it, but does not think it will work. And therefore it does not fully work, because you... it's hard. I think it's really hard to have one leg in it, and one outside, you know...'' [Scrum Master driving the agile transformation]
\end{quote}

From the hardware-focused participants, they described their agile implementation as something that needed to be agile in itself. That means that they had to related their new process to also fit with the old, which was described as challenging.

\begin{quote} 
``We have tried to see the work process as something agile and adaptable too! Now there are more clear toolboxes with how to work in an agile way in the company. When we started that didn't exist, but, I mean that, when you talk about Scrum as an agile method it's not anything more than a toolbox with different practices, and you have to try to pick what suits your organization and your... we started with that mindset that we would have `what is the smallest part of scrum we can pick?' or `what is the smallest set of tools we can make use of?' and then, I guess, with time we have realized that, yes, `we need this' or 'we need do do that one too'. [Scrum Master driving the agile transformation]
\end{quote}

To summarize, the participants' leadership style is also adapted to where the company is in its agile journey, not just the internal process structure, as in the previous theme. The leaders act as both more traditional managers and as more agile (i.e. contextually adapted) leaders depending on what is needed at that point in time.

\section{Discussion}\label{sec:discussion}
This study found that the main challenges in relation to group dynamics in the agile context is to adapt the leadership to the (1) teams' collaborative maturity, (2) design new teams well, and (3) balance the ``old'' ways of working with the intended new agile processes and their innate different culture. The first one, called (1) \emph{Team maturity} included that the agile leaders saw a need to step back from mature teams and then be the facilitator with a strong focus on impediments external to the team. However, as agile leaders, the challenge from their perspective was that newly formed, or less mature teams, on the contrary, needed a lot of support in order to grow into agile teams. This was not seen a ``agile'' by the interviewees and expressed as frustrating.  

The second category (2) \emph{Team design} is connected to the first category but focuses on that teams need to be well designed in the agile context and get much more help setting up the agile team than the interviewees initially had thought. Teams do not seem to be able to design themselves when starting their agile journey, but instead need a suggested initial team design. New teams, that are less mature by definition, cannot tailor their own agile process, but need help to get started and can then improve their process based on what they learn about their ecosystem. The third category (3) \emph{Culture and mindset} is about that the needed adaptability of agile teams are also in relation to where the whole company is in its overall agile journey. Agile leaders described that they acted as more traditional managers sometimes, if needed, but also adapted to where in the company teams were more allowed to drive themselves within their predefined mandate.

That team maturity is connected to team agility has been shown in previous studies (e.g.\ Gren et al.\ \cite{gren2019agile}). The third category also confirms previous results on the difficulty with integrating an agile approach to a traditional context (e.g.\ \cite{hodgson2013controlling}). The second category, though, on the importance of team design, as conducted by a leader more external to the team, seems to be a novel finding. Even with regards to the other two categories, this study adds to knowledge since it focuses on the appointed agile leaders and what challenges they define in relation to building agile teams. More immature agile teams were said to be more open to all the practices, and that agile leaders need to provide a clear agile work process to the teams for them to get started in a constructive way. While this is not according to the description of an agile and self-organizing team, this is a typical trait of groups in the Forming stage according to group development \cite{tuckman}. Team members are focused on dependency and inclusion when newly formed, and need to build some initial trust before sharing opinions and questioning each other. Giving new teams a structure and directives would therefore help them in their development as compared to trying to get them to self-organize too early. In the results of this study, we clearly see that the agile leaders describe a need for more guidance of newly formed agile teams as well. Not only do teams need more guidance in their internal collaboration, but just like Wageman \cite{wageman2001leaders} concluded in her studies on self-organizing teams in general, someone needs to design the teams well also before they are deployed. This aspect seems to be largely ignored in the agile literature. After the team is set up, the role of the leader reassembles the leadership needed in different group development stages, which verifies their importance and clarified the connection between the two concepts. This study has shown that this temporal perspective of needed leadership when creating self-organizing teams is also essential when setting up agile teams, and Wageman's \cite{wageman2001leaders} theory fits the result of this study exceptionally well. \emph{If we aim for having only a facilitating and coaching leadership from day one, we hamper the teams' development instead of the opposite.}

The interviewees also described the more mature agile teams as aware of their context, e.g.\ surrounding teams and overall company strategies, and having the ability to tailor their own process in relation to responding to a change in requirements. This is at the very core of agility \cite{scrumorbeing}, and shows that self-organization is a property of collaboratively mature teams. In contrast, one interviewee thought that following the agile process was bad and thus that an agile process was inflexible. Adapting agile methods to a large organization was described as a balance between old and the new work methods, and therefore, there is a large risk of adapting the old organization to the new agile structure and by this not perform an actual agile transformation but a renaming of the existing structure. One example would be to keep a command-and-control approach by the line management instead of providing a structure for teams to grow and eventually self-organize, i.e.\ letting team members lead when they are ready to take on and share the leadership function more. This challenge was more stressed by our two participants from hardware development. A key to create agile teams, as described by our interviewees, seems to be to suggest good work structures for teams and then let them adjust their process as the development cycle moves along. This is a tricky balance for new agile leaders since new teams need more structure but the appointed leaders then need to take a step back and let the team self-organize when ready. Therefore, a mix of more directive leadership styles in combination with consultative and coaching leadership styles, seems to be what the successful practitioners do in practice. This is not then to be confused with traditional line management control, but instead the application of different types of support for teams guided by the end-goal of the team leading itself.

The guidelines from the agile community have been to be a facilitator of the process only instead of being a more directive leader \cite{rising2000scrum}. The problem with such simplified guidance to new leaders is that such behavior only works in very special cases. With a mature team, and with good organizational support, taking a step back, delegate and be a process facilitator is easy. The problem is that becoming a leader in such a context immediately is pretty rare. There is an awareness in practice that inexperienced teams need more guidance, but this awareness is only in relation to agile practices \cite{adkins} and not group dynamics over time. Therefore, that explanation model of agility fails to explain why some teams do not become agile over time even if they learn the agile practices. An understanding of group developmental psychology provides that explanation \cite{tuckman}, and explains why different leadership styles and team support are needed at the different stages. And maybe even more novel in this current study, teams need to be designed well before they are deployed if they are to self-organize as fast as possible \cite{wageman2001leaders}. This motivates the future inclusion of more group dynamics teaching in agile courses both at companies and at universities.

\section{Threats to validity}\label{sec:limits}
Since leadership is a complex construct, a qualitative method was suggested to provide deeper insights of the reality. However, with fewer participants is becomes more difficult to generalize to a larger population, which should be done with care for our study. The selection of a large variety of interviewees from various organizational settings in the sample, is one major strength of the study, but conclusions from our two participants from hardware development should of course be drawn with even more care and only seen a small additional comparison. The diverse sample with regards to companies provides a broad view of experiences from practitioners working with an agile approach in different stages of its implementation. A high risk in qualitative studies is that our interpretations of what the interviewees said could of course be erroneous and prone to confirmation bias. A way to counter such a threat was to be transparent in the description of our research method, and to be transparent with what quotes we believe supported our claims. In the analysis, we acknowledged our previous knowledge of group dynamics, which many of our participants did not have. Our participants could have a different perception of the constructs under investigation, but as researchers from different areas, one with and one without previous knowledge of agile but both with rigorous experience of leadership, we believe that this combination was advantageous for the analysis. For validation of interpretations and citations, feedback from the interviewees might have improved the quality of the analysis. We recommend that procedure for upcoming studies in the area. We also acknowledge that having a mix of two languages (English and Swedish), and conducting interviews in English with non-native speakers also threatens the validity of this study due to difficulty in obtaining exact translations between languages.

\section{Conclusion and Future Work}\label{sec:limits}
This paper set out to investigate what challenges appointed agile leaders see in relation to group dynamics. Through a qualitative method of interviews and a thematic analysis, we have found that leader adaptability to team maturity, the careful design of new teams, and a continuous balance between traditional and new work principles are all essential to the success of an agile transformation. These findings are an important contribution to both research and practice since it gives an in-depth view of leadership challenges of group dynamics in agile transformations at different scales. In terms of future research, we particularly suggest more use of qualitative research methods both when studying leadership in engineering, but also to apply narrative analysis \cite{smith2006narrative} to large agile transformations, since analyzing the complete transition from a waterfall process to an agile approach has not been done in its entirety. We believe the narrative analysis method is underused in software engineering research and fits well to the study of organizational changes over time.

\section*{Acknowledgment}\label{sec:limits}
We would like to thank all the participants and everyone who helped us with making this study possible, and we would like to acknowledge Jennifer Strand and Petra Bostr{\"o}m specifically for their excellent support.
\bibliographystyle{splncs}
\bibliography{ref}
\newpage

\end{document}